# CIGI QUALITA MOSIM 2023
# L'origine de l'objectif est-elle importante? Effets motivationnels d'objectifs autodéfinis en production

Mario Passalacqua[1], Robert Pellerin[1], Florian Magnani[2], Laurent Joblot[3]
Esma Yahia[4], Frédéric Rosin[4] et Pierre-Majorique Léger[5]

[1] Département de mathématiques et de génie industriel, Polytechnique Montréal
2500 Chem. de Polytechnique, Montréal, QC H3T 1J4, Canada
mario.passalacqua.nom@polymtl.ca ; robert.pellerin@polymtl.ca

[2] CERGAM et DynEO, École Centrale de Marseille
38 Rue Frédéric Joliot Curie, 13013 Marseille, France
Florian.magnani@centrale-marseille.fr

[3] Arts et Métiers institute of technology, LISPEN / HESAM University / UBFC
11 Rue Georges Maugey, 71100 Chalon-sur-Saône, France
Laurent.joblot@ensam.eu

[4] LISPEN et DynEO, Arts et Métiers ParisTech
2 Cr des Arts et Métiers, 13617 Aix-en-Provence, France
Esma.Yahia@ensam.eu ; Frederic.Rosin@ensam.eu

[5] Département de technologies de l'information, HEC Montréal
3000 Chem. de la Côte-Sainte-Catherine, Montréal, QC H3T 2A7, Canada
pierrre-majorique.leger@hec.ca

*Résumé* - Seuls 21 % des employés se considèrent engagés au travail. Le désengagement, générant une variété de résultats négatifs dont l'absentéisme et la perte de productivité, est encore plus problématique lorsque le travail est de nature répétitive. La gamification, c'est-à-dire l'intégration d'éléments de jeu dans les systèmes de travail, est une solution possible permettant d'accroître l'engagement et la motivation. Dans la présente étude, nous nous concentrons sur un de ses leviers, soit la fixation d'objectifs. Nous soutenons que les objectifs assignés produisent une motivation extrinsèque qui n'améliore l'engagement et la performance qu'à court terme. Nous postulons aussi que les objectifs autodéfinis conduisent à une motivation autodéterminée, et donc, à un engagement et une performance accrue à long terme. Ainsi, un cadre expérimental impliquant cent deux participants effectuant une tâche répétitive dans une de trois conditions (objectif assigné, autodéfini, aucun objectif) a été réalisé. Les résultats ont montré que l'autonomie perçue et la performance étaient meilleures lorsque les objectifs étaient autodéfinis. Cependant, l'engagement demeure égal lorsque les objectifs sont autodéfinis et assignés. Ces résultats suggèrent que les objectifs autodéfinis ont un plus grand potentiel de générer des résultats positifs à long terme.

*Abstract* - Only 21% of employees consider themselves engaged at work. Moreover, disengagement has been shown to be even more problematic when work is repetitive in nature. Lack of engagement has been linked to variety of negative outcomes for employees and companies (e.g., turnover, absenteeism, well-being, safety incidents, productivity). Gamification, i.e., integrating game elements into work systems, has been successfully used to increase engagement and motivation, even when work tasks were mundane and repetitive. In the current study, we focused on a commonly used game element, goal setting, through the lens of self-determination theory and goal-setting theory. We argue that goals given by an external source (e.g., company, experimenter) produce extrinsic motivation, which improves engagement and performance only in the short term. We posit that self-set goals lead to more autonomous motivation, and therefore long-term engagement and performance. One hundred two participants completed a repetitive material-handling task in one of three conditions (assigned goal, self-set goal, no goal). Results showed that perceived autonomy (autonomous motivation) and performance were best when goals were self-set. Engagement, however, was equal between self-set and assigned goals. The results indicate that self-set goals have the greatest potential to generate long-term positive outcomes both for employees and companies.

*Mots clés* - Motivation, Engagement, Gamification, Manutention de matériel, Théorie de l'autodétermination

# 1 INTRODUCTION

Seuls 21% des employés dans le monde se sentent engagés dans leur travail (Gallup, 2023). Cette statistique est inquiétante si l'on considère qu'une récente méta-analyse de 339 études (portant sur 230 organisations dans 49 secteurs d'activité de 73 pays) a révélé qu'un manque d'engagement des employés entraîne une baisse de la performance de l'entreprise dans neuf volets distincts : fidélité des clients, rentabilité, productivité, chiffre d'affaires, incidents de sécurité, perte de stock, absentéisme, incidents de sécurité et qualité des produits (Harter et al., 2016). L'engagement des employés devient encore plus problématique pour les tâches qui sont de nature répétitive (Thomas et Holley, 2012 ; Tims et Bakker, 2013). Dans ce contexte spécifique, un manque d'engagement des employés se retrouve corrélé à une diminution de la rentabilité de l'entreprise, à une augmentation des défauts, à une augmentation de l'incidence de la sécurité, à une augmentation du chiffre d'affaires et à une diminution de la productivité (Manhattan Associates, 2017 ; Santhanam & Srinivas, 2019).

Une solution possible pour accroître l'engagement des employés est la gamification. La gamification consiste à appliquer la mécanique des jeux à des contextes non ludiques, tels que les tâches professionnelles, afin d'accroître la motivation (Deterding et al., 2011). Les éléments courants de la gamification sont la fixation d'objectifs (par exemple, le temps), les tableaux de classement, les écussons et les points. La gamification peut rendre les tâches répétitives ou banales plus intéressantes et stimulantes. En effet, la recherche a montré que la gamification peut être efficace pour augmenter la motivation et l'engagement des employés et améliorer la performance globale dans divers environnements de travail, notamment pour des tâches répétitives (Bahr et al., 2022 ; Klevers et al., 2016 ; Passalacqua, Léger, et al., 2020 ; Passalacqua, Sénécal, et al., 2020 ; Small, 2017). Cependant, les éléments de jeu dans les systèmes gamifiés augmentent prioritairement la motivation extrinsèque, qui est associée à des augmentations à court terme de l'engagement et de la performance (Duggal et al., 2021 ; Lamprinou & Paraskeva, 2015 ; Mekler et al., 2017). Les éléments de jeu qui génèrent de la motivation autodéterminée/intrinsèque, en revanche, sont associés à un engagement à long terme et ont donc un impact durable sur la performance des employés et de nombreux autres résultats positifs tels que le comportement de citoyenneté organisationnelle et la performance organisationnelle (Deci et al., 2017b ; Meyer & Gagne, 2008 ; Van den Broeck et al., 2021).

Dans la présente étude, nous avons cherché à approfondir l'un des éléments de jeu les plus courants, la fixation d'objectifs de temps. Nous voulions disséquer l'effet de la source des objectifs sur la motivation autodéterminée/extrinsèque, l'engagement et la performance des participants. Plus précisément, nous cherchons à comparer comment ces variables sont influencées lorsque les objectifs temporels sont fixés par les participants eux-mêmes ou par une personne extérieure (c'est-à-dire l'expérimentateur). La recherche a montré que le fait de fixer son propre objectif peut augmenter la motivation autodéterminée, l'engagement et la performance, conduisant ainsi à une variété de meilleurs résultats à long terme (Erez & Arad, 1986 ; Harkins & Lowe, 2000 ; Passalacqua, Léger, et al., 2020 ; Welsh et al., 2020). À notre connaissance, une seule étude a testé expérimentalement l'influence de la source de fixation d'un objectif dans un contexte de gamification (Passalacqua, Léger, et al., 2020). Cette étude a révélé que les objectifs fixés par soi-même conduisaient au plus grand potentiel de motivation autodéterminée, d'engagement et de performance à long terme dans un contexte d'entreposage. Dans le but d'étendre cette recherche au milieu de la production, nous avons mené une expérience dans laquelle 102 participants ont effectué une tâche répétitive de démontage de matériel. Les participants ont soit choisi leur propre objectif de temps, soit se sont vu assigner un objectif, soit n'avaient pas d'objectif (groupe contrôle).

La suite de l'article est structurée de la manière suivante : tout d'abord, nous passons en revue la littérature pertinente et développons nos hypothèses ; ensuite, nous présentons notre méthodologie; troisièmement, nous présentons nos résultats; et enfin, nous discutons ces derniers.

# 2 REVUE DE LA LITTÉRATURE ET DÉVELOPPEMENT D'HYPOTHÈSES

La section suivante passe en revue la littérature pertinente sur l'engagement des employés, ses antécédents et la théorie de fixation des objectifs (goal-setting theory).

## 2.1 Engagement des travailleurs

L'engagement dans la tâche est un concept multidimensionnel composé (1) d'une dimension d'état psychologique, (2) d'une dimension comportementale et (3) d'une dimension dispositionnelle (trait de personnalité) (Macey & Schneider, 2008 ; Meyer et al., 2010). La sous-dimension d'état psychologique est composée d'une composante cognitive et d'une composante émotionnelle (Kahn, 1990 ; Schaufeli et al., 2002). L'engagement cognitif se caractérise par une concentration totale et une absorption mentale dans une tâche. L'engagement émotionnel englobe l'émotion positive et négative (valence) et l'activation émotionnelle (Lang, 1995 ; Macey & Schneider, 2008). L'engagement comportemental se réfère aux indicateurs observables de l'engagement dans une tâche de travail (Macey & Schneider, 2008). L'engagement dispositionnel est défini comme la prédisposition d'un travailleur à vivre son travail de manière positive, active et énergique et à se comporter de manière adaptée (Macey & Schneider, 2008). Cela signifie que certains travailleurs sont plus susceptibles que d'autres de percevoir, de se comporter et de penser d'une manière qui leur permettra d'être motivés ou engagés (Deci & Ryan, 1985 ; Meyer et al., 2010). Les traits de personnalité d'un travailleur ont essentiellement un impact sur la manière dont il évalue cognitivement une situation comme étant contrôlante ou favorisant l'autonomie, ce qui a une incidence sur la motivation autodéterminée ou extrinsèque et, par conséquent, sur l'engagement (Ryan & Deci, 2008 ; Szalma, 2020).

## 2.2 Antécédents de l'engagement (théorie de l'autodétermination)

Les antécédents de l'engagement des employés, c'est-à-dire la motivation autodéterminée et la satisfaction des besoins, peuvent être mieux compris à travers le prisme de la théorie de l'autodétermination (TAD). La TAD est une théorie issue de la psychologie de la motivation humaine qui a été affinée au cours de quatre décennies et utilisée dans une variété de contextes. Elle est devenue la principale théorie permettant d'expliquer l'effet motivationnel des jeux (Deci et Ryan, 1980 ; Ryan et al., 2006). À la base, la TAD explique et prédit comment les situations, les contextes ou les événements affectent la motivation d'une personne (Deci et al., 2017). Dans un contexte de tâches répétitives, où la monotonie et l'ennui affectent directement la motivation et la performance des travailleurs, la TAD peut être particulièrement utile en fournissant des indications exploitables pour atténuer les effets néfastes sur la motivation et la performance (Small, 2017).

La motivation se situe sur un continuum, allant de l'intrinsèque à une extrémité, à l'extrinsèque au centre et à l'amotivation à l'autre extrémité. La motivation intrinsèque est liée à l'exécution d'une action pour son plaisir intrinsèque et parce qu'elle correspond aux valeurs, aux intérêts ou aux aspirations de l'individu. La motivation intrinsèque représente le type le plus autodéterminé et est le plus fort prédicteur du bien-être des travailleurs et de l'absentéisme (Van den Broeck et al., 2021). La motivation extrinsèque concerne l'exécution d'une tâche en raison d'une demande externe, comme une récompense externe ou l'évitement d'une punition. L'amotivation représente un manque total de motivation pour réaliser une action. Une motivation plus régulée intrinsèquement conduira les travailleurs à être plus engagés, innovants, généralement plus heureux, moins susceptibles de changer d'emploi et à mieux accepter la technologie (Deci et al., 2017 ; Meyer & Gagne, 2008 ; Venkatesh et al., 2002). Lorsque ces besoins sont contrariés, les travailleurs sont moins intrinsèquement motivés, ce qui est associé à l'épuisement professionnel, au stress, au désengagement et d'autres facteurs qui affectent leur bien-être (Deci et al., 2017).

La principale prémisse de la TAD est que la satisfaction des besoins psychologiques d'autonomie, de compétence et de relation des travailleurs conduit à une plus grande internalisation du motif d'exécution d'une tâche professionnelle, c'est-à-dire le degré auquel le motif d'exécution d'une action est accepté et intégré en soi. L'autonomie (1) fait référence au sentiment que le comportement ou les actions viennent de soi-même, par opposition aux facteurs externes ; la compétence (2) fait référence au sentiment d'avoir un effet sur notre environnement ; l'appartenance sociale (3) fait référence au sentiment d'avoir des interactions significatives avec les autres (Ryan & Deci, 2000).

Tout comme les comportements ou les actions, les raisons de poursuivre des buts sont internalisées à des degrés divers et produisent donc différents types de motivation le long du continuum d'autodétermination. De nombreuses recherches ont montré que les buts qui sont intériorisés dans une plus large mesure donnent lieu à une plus grande réussite des objectifs (Koestner & Hope, 2014). En d'autres termes, lorsque la motivation sous-jacente à la poursuite d'un but se rapproche davantage du type intrinsèque, les buts sont plus souvent atteints, car les individus fournissent plus d'efforts et rencontrent moins de conflits. Puisque la satisfaction des trois besoins psychologiques fondamentaux favorise l'intériorisation et une motivation plus autonome, la TAD postule que ces besoins doivent être pris en compte lors de la fixation des objectifs. Cela implique que l'autonomie, la compétence et l'appartenance sociale sont essentielles à la réussite de la fixation et de la poursuite des objectifs.

### 2.3 Goal-setting theory (théorie de fixation des objectifs)

Contrairement à la TAD, qui se concentre sur les raisons sous-jacentes de la poursuite des objectifs, la théorie de la fixation des objectifs se concentre principalement sur la meilleure façon de fixer des objectifs. La théorie de la fixation des objectifs, qui a émergé de centaines de résultats empiriques, affirme que la fixation des objectifs est liée à la performance (Locke & Latham, 1990). Au fur et à mesure que la recherche progresse, il semble qu'il y ait maintenant un consensus sur les caractéristiques des objectifs qui conduisent aux meilleures performances. Les objectifs doivent être spécifiques, la progression de l'objectif doit être mesurable, les objectifs doivent être réalisables grâce à un effort personnel, la difficulté de l'objectif doit être réaliste et les objectifs doivent être limités dans le temps. La théorie de la fixation des objectifs fait également la distinction entre les objectifs fixés par soi-même et les objectifs assignés. Les résultats ont montré que les objectifs fixés par soi-même semblent être plus efficaces dans la poursuite à long terme des objectifs (Erez & Arad, 1986 ; Harkins & Lowe, 2000 ; Locke & Latham, 2002 ; Passalacqua et al., 2020 ; Welsh et al., 2020). Les objectifs choisis par soi-même satisfont mieux l'un des trois moteurs de la motivation autodéterminée (l'autonomie), ce qui conduit à un plus grand engagement et à de meilleures performances.

Par conséquent, nous supposons que les objectifs que l'on se fixe soi-même donnent de meilleurs résultats qu'en cas d'objectifs assignés ou d'absence d'objectifs.

- H1 : Les objectifs choisis par soi-même amèneront à une plus grande satisfaction des besoins (autonomie) qu'en cas d'objectifs assignés ou d'absence d'objectifs.
- H2 : Les objectifs choisis par soi-même amèneront à un plus grand engagement dans la tâche qu'en cas d'objectifs assignés ou d'absence d'objectifs.
- H3 : Les objectifs choisis par soi-même amèneront à une meilleure performance qu'en cas d'objectifs assignés ou d'absence d'objectifs.

## 3 MÉTHODOLOGIE

La section suivante présente le plan expérimental, l'échantillon, la tâche, la procédure, l'opérationnalisation des variables, les mesures et l'analyse statistique de notre étude.

### 3.1 Plan expérimental et échantillon

Au total, 102 participants ont participé à une expérience en laboratoire (67 hommes, 35 femmes). L'âge moyen des participants était de 21,97 ans (écart-type = 2,69). Les participants ont été recrutés dans deux universités françaises. Aucun participant n'avait d'expérience préalable de la tâche choisie. Cette expérience a été examinée et approuvée par le comité d'éthique de la recherche de notre institution (certificat #2023-5058). Chaque participant a reçu 40€ pour sa participation à la fin de l'expérience. L'expérience a été menée en français. Cette étude a utilisé un plan d'étude inter-sujet. La source de fixation des objectifs a été manipulée, ce qui a donné lieu à trois conditions : (1) objectifs autodéfinis (SSG) ; (2) objectifs assignés (AG) ; et aucun objectif (NG). Les participants ont été assignés au hasard à l'une de ces conditions.

### 3.2 Tâche et configuration expérimentale

Les participants ont été informés qu'ils étaient le premier opérateur sur une ligne de démontage de raquettes à neige. Leur objectif était de désassembler une partie (environ 10%) de la raquette à neige le plus rapidement possible. La tâche consistait à démonter 60 raquettes à neige, qui étaient placées sur des supports (120cm x 71cm x 103cm) à côté du participant. En clair, les participants prenaient une raquette sur le support, en démontaient le premier 10%, puis la remettaient sur le support. Cette tâche a été choisie pour simuler une tâche simple, monotone et désengageante de manipulation et de désassemblage de matériel. Le poste de travail du participant est illustré à la Figure 1.

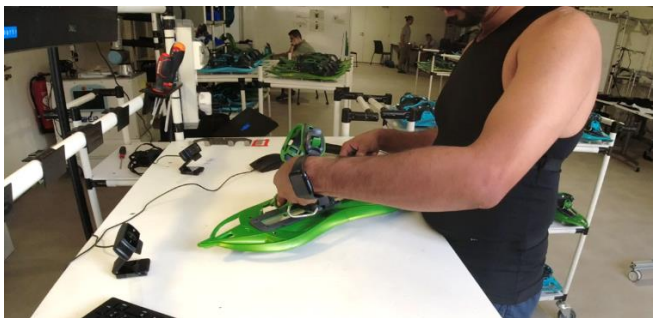
**Figure 1. Poste de travail du participant**

### 3.3 Procédure

Les participants ont été informés que l'expérience consistait à évaluer leur capacité à démonter des raquettes à neige. Après avoir signé le formulaire de consentement, les instruments physiologiques ont été installés sur les participants. Ils ont ensuite suivi une formation, au cours de laquelle la tâche a été expliquée et démontrée. Dans le cadre de cette formation, les participants ont également effectué le démontage de trois raquettes à neige, afin de se familiariser avec la tâche. Les participants ont ensuite répondu à un questionnaire de pré-expérience comprenant des données démographiques et un questionnaire de personnalité (engagement dispositionnel) avant de passer à la tâche expérimentale. Les participants ont ensuite effectué la tâche dans l'une des trois conditions. Dans toutes les conditions, les participants devaient accomplir la tâche le plus rapidement possible.

Dans la condition de l'objectif autodéfini, les participants devaient choisir l'un des trois objectifs de temps (17 min 33 s, 13 min 52 s, ou 10 min 11 s pour le lot de 60 raquettes). Ces temps ont été calculés à partir d'un test pilote réalisé au préalable auprès de 10 participants. 17 min 33 s représente le temps moyen de démontage, 13 min 52 s représente le temps moyen moins un écart-type, et 10 min 11 s représente le temps moyen moins deux écarts-types. Les participants ont été informés de ces données au moment de prendre leur décision. Ils ont également été informés que tous les temps étaient réalisables, mais que 10 min 11 s était très difficile et 13 min 52 s était difficile. Dans la condition de l'objectif assigné, tous les participants se sont vu attribuer l'objectif intermédiaire (13 min 52 s). Ils ont également été informés que cet objectif était réalisable, mais difficile. Dans la condition sans objectif, les participants n'avaient pas d'objectif de temps.

### 3.4 Opérationnalisation et mesures des variables

Le Tableau 1 présente un résumé de l'opérationnalisation des variables et des mesures. Lorsque cela était possible, les variables et construits ont été évalués en utilisant une approche multiméthode (perceptive et physiologique). Les mesures physiologiques nous ont permis de mesurer l'état d'un participant, sans interruption, pendant toute la durée de la tâche, limitant ainsi les biais associés à l'utilisation de seulement des mesures perceptives (de Guinea et al., 2014).

#### 3.4.1 Temps de performance

La performance a été opérationnalisée comme le temps nécessaire à un participant pour accomplir la tâche. Un temps plus bas indique une meilleure performance.

#### 3.4.2 Autonomie perçue

L'autonomie perçue a été mesurée à l'aide de la sous-échelle autonomie (autodétermination) du « Psychological empowerment scale » (Spreitzer, 1995). Cette sous-échelle est composée de 3 items sur une échelle de Likert en cinq points. Les autres besoins psychologiques, la compétence et l'appartenance sociale n'ont pas été abordés dans cette étude, car la source de fixation de l'objectif ne devrait pas avoir d'impact sur la perception de la compétence ou d'appartenance sociale des participants.

#### 3.4.3 Engagement cognitif dans une tâche

##### 3.4.3.1 Autorapporté

L'engagement cognitif a été mesuré à l'aide de la sous-échelle d'absorption de l'échelle d'engagement au travail d'Utrecht (Schaufeli et al., 2003). Nous avons utilisé la version française du questionnaire (Zecca et al., 2015). Ce questionnaire est composé de 9 items notés sur une échelle de Likert en sept points.

##### 3.4.3.2 Physiologique

Le gilet intelligent Hexoskin (Carré Technologies, Montréal, Canada) a été utilisé pour enregistrer les données de fréquence cardiaque et de respiration. Ce gilet capturait les données de l'électrocardiogramme à une dérivation de 256 Hz à l'aide d'une électrode intégrée, les données de la respiration de 128 Hz à l'aide de deux capteurs pléthysmographiques respiratoires inductifs intégrés et les données d'accélération/activité de 64 Hz à l'aide d'un accéléromètre à trois axes intégré. L'utilisation du gilet intelligent Hexoskin a été évaluée et validée par une multitude d'études (Cherif et al., 2018 ; Jayasekera et al., 2021) En utilisant une transformation de Fourier rapide sur les intervalles entre les battements (intervalles RR), nous avons dérivé la puissance absolue pour les bandes de basse fréquence (LF) 0,04-0,15 Hz) et de haute fréquence (HF) (0,15-0,4 Hz) de la variabilité du rythme cardiaque (VRC) des participants. La bande LF de la VRC est produite à la fois par le système nerveux parasympathique (SNP) et le système nerveux sympathique (SNS). En revanche, il a été démontré que la bande HF de la VRC est produite principalement par le SNS (Shaffer et al., 2014). Le SNS contrôle principalement les réponses « fight or flight », tandis que le SNP contrôle principalement les réponses « rest and digest ». Le rapport LF/HF a pour but d'estimer le rapport entre l'activité du SNS et du SNP et il a été démontré qu'il était un indicateur de l'engagement cognitif pendant une tâche (Gao et al., 2020).

#### 3.4.4 Engagement émotionnel dans la tâche

##### 3.4.4.1 Autorapporté

La composante émotionnelle de l'engagement dans la tâche d'état peut être subdivisée en deux composantes : la valence (bonheur/tristesse) et l'activation (intérêt/ennui) (Lang, 1995 ; Macey & Schneider, 2008). Nous avons utilisé le curseur affectif de Betella et Verschure (2016), qui se compose de deux curseurs mesurant la valence et l'activation sur une échelle de 0 à 100. Le curseur de valence a la tristesse à une extrémité (0) et le bonheur (100) à l'autre, tandis que le curseur d'activation a l'ennui (0) à une extrémité et l'intérêt (100) à l'autre.

##### 3.4.4.2 Physiologique

Nous avons utilisé la fréquence respiratoire comme indicateur de l'activation émotionnel. Cette mesure a été validée par Bradley et Lang (2007). La fréquence respiratoire a été mesurée à l'aide des deux capteurs pléthysmographiques inductifs respiratoires intégrés au gilet Hexoskin. Les données relatives à la respiration ont été normalisées au niveau du participant en utilisant une période d'inactivité debout comme référence.

#### 3.4.5 Engagement comportemental dans la tâche

##### 3.4.5.1 Autorapporté

L'engagement comportemental a été mesuré à l'aide de la sous-échelle de vigueur de la version française de l'échelle d'engagement au travail d'Utrecht (Schaufeli et al., 2003).

### 3.4.5.2 Physiologique

L'engagement comportemental est opérationnalisé comme l'écart type de l'intensité de l'effort physique manifesté pendant une tâche, mesuré en g-force (Gao et al., 2020). Cette mesure est captée par l'accéléromètre à 3 axes du gilet Hexoskin.

### 3.4.6 Engagement dispositionnel

L'engagement dans la tâche a été mesuré à l'aide de la version française de l'échelle d'orientation générale vers la causalité (Deci & Ryan, 1985 ; Meyer et al., 2010). La version française a été validée par Vallerand et al. (1987). Elle consiste en 12 vignettes décrivant une situation d'orientation vers l'accomplissement, à propos de laquelle le participant doit répondre à trois questions pour chacune d'entre elles en utilisant une échelle de Likert en sept points. Chacune des trois questions représente une sous-échelle du questionnaire : autonomie, contrôle et impersonnel. Les participants qui obtiennent un score élevé en autonomie sont plus susceptibles de percevoir les situations ou les tâches comme favorisant l'autonomie et sont donc plus susceptibles de ressentir une motivation autodéterminée et un engagement plus élevé. Les participants ayant un score plus élevé en contrôle sont plus susceptibles de percevoir les situations ou les tâches comme étant contrôlées par une source externe et sont donc plus susceptibles de ressentir une motivation extrinsèque et un engagement moindre par rapport à ceux qui ont un score plus élevé en autonomie. Les participants ayant un score plus élevé en impersonnel sont plus susceptibles de se sentir incapables d'avoir un effet ou de contrôler les situations ou les tâches. Ils peuvent éprouver un sentiment d'impuissance et sont plus susceptibles d'éprouver de l'amotivation et un manque d'engagement (Deci et al., 2017 ; Ryan & Deci, 2008 ; Szalma, 2020). Ces différences interpersonnelles concernant les prédispositions à éprouver certains types de motivation et un certain niveau d'engagement introduisent une variabilité intragroupe dans nos conditions expérimentales qui peut influencer la façon dont la fixation des objectifs (variable indépendante) affecte nos variables dépendantes. Par conséquent, nous avons tenté de réduire cet effet en l'introduisant comme covariable dans notre modèle statistique.

**Tableau 1 : Opérationnalisation des mesures et des variables**

| Variable | Mesure | Type de mesure | Opérationnalisation |
|---|---|---|---|
| Perception d'autonomie | Psychological empowerment scale | Auto-évaluation | Score d'autonomie |
| Engagement cognitif | Sous-échelle d'absorption de l'UWES | Auto-évaluation | Score d'absorption cognitive |
| | Veste Hexoskin | Physiologique | Rapport LF/HF |
| Engagement émotionnel (activation) | Curseur affectif | Auto-évaluation | Score d'activation émotionnelle |
| | Veste hexoskin | Physiologique | Taux de respiration |
| Engagement émotionnel (valence) | Curseur affectif | Auto-évaluation | Score de valence émotionnelle |
| Engagement comportemental | Sous-échelle de la vigueur de l'UWES | Auto-évaluation | Score de vigueur |
| | Veste Hexoskin (accéléromètre) | Physiologique | Écart-type de l'intensité de l'effort physique fourni lors d'une tâche (mesuré en g-force) |
| Engagement dispositionnel | Échelle d'orientation générale de la causalité | Auto-évaluation | Autonomie, contrôle et score impersonnel |
| Temps d'exécution de la tâche | Enregistrement vidéo | Observation | Temps mis par un participant pour accomplir la tâche (plus bas = meilleur) |

### 3.5 Analyse statistique

Une analyse de variance (ANOVA) de type 3 a été réalisée pour déterminer l'effet global de la source de fixation de l'objectif sur chacune des variables dépendantes, en contrôlant pour l'engagement dispositionnel (covariable). Lorsque nous avons trouvé des effets globalement significatifs, nous avons utilisé des régressions linéaires pour comparer les moyennes des moindres carrés par paire. Ces tests ont été ajustés pour les comparaisons multiples en utilisant la méthode de Holm. Nous avons utilisé R (langage de programmation) et la version 26 de Statistical Product and Service Solutions (SPSS) pour effectuer nos analyses.

### 3.6 Calcul de la puissance statistique a priori

Les calculs de puissance statistique a priori permettent d'estimer la taille de l'échantillon nécessaire pour atteindre un niveau statistique significatif. La puissance statistique fait référence à la probabilité de détecter correctement des différences au sein de l'échantillon (Cohen, 1992). Le logiciel G*Power (Faul et al., 2009) a été utilisé pour calculer la puissance dans la phase de planification de l'expérience avec les paramètres suivants. La taille de l'effet a été estimée sur la base d'une étude similaire (Passalacqua, Léger, et al., 2020). Nous avons calculé la moyenne de toutes les tailles d'effet trouvées dans cette étude, ce qui a conduit à une taille d'effet approximativement comprise entre moyenne et forte. Nous avons donc utilisé une taille d'effet moyenne-forte (f=0,32). Une puissance de 0,80 a été choisie, comme recommandé (Cohen, 1992). En bref, les calculs de puissance nous ont appris qu'un échantillon de 98 participants était jugé suffisant pour rejeter correctement l'hypothèse nulle avec une certitude de 80 %.

## 4 RÉSULTATS

Deux participants ont été exclus en raison d'une défaillance de l'équipement. Les résultats ont montré que la difficulté du but choisi (moyenne, 1 écart-type sous la moyenne ou 2 écarts-types sous la moyenne) n'avait aucun effet sur les variables dépendantes.

### 4.1 H1 : Les objectifs choisis par soi-même amèneront à une plus grande satisfaction des besoins (autonomie) comparativement aux objectifs assignés ou l'absence d'objectifs.

Une ANOVA de type 3 a montré un effet principal significatif de la source des objectifs sur l'autonomie après contrôle de l'engagement dispositionnel, $F(2, 91) = 4,36$, $p = 0,016$, $\eta p2 = 0,09$. Des régressions linéaires par paire post-hoc ont révélé que l'autonomie était significativement plus importante lorsque les objectifs étaient autodéfinis que lorsqu'ils étaient assignés ($t = 2,75$, $p = 0,018$). Cependant, aucune différence n'a été observée entre autodéfini et absence de but ($t = 2,138$, $p = $ ) ou entre assigné et l'absence de but ($t = -0,61$, $p = 1$). La Figure 2 illustre ces résultats.

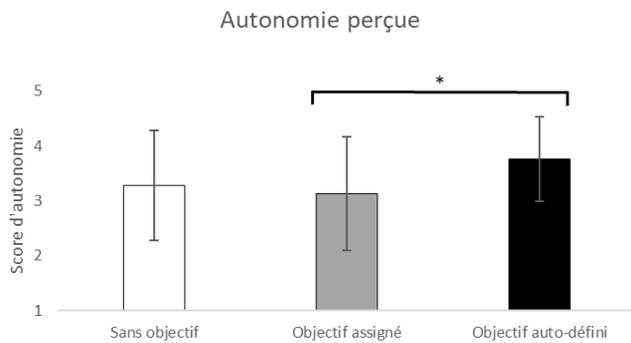

**Figure 2. Autonomie perçue par condition**

*4.2 H2 : Les objectifs choisis par soi-même amèneront à un plus grand engagement dans la tâche comparativement aux objectifs assignés ou l'absence d'objectifs.*

### 4.2.1 Engagement cognitif (état)
#### 4.2.1.1 Auto-évaluation
Une ANOVA de type 3 n'a montré aucun effet principal significatif de la source de l'objectif sur l'absorption autorapportée après contrôle de l'engagement dispositionnel, $F(2, 91) = 2,59$, $p = 0,081$.

#### 4.2.1.2 Physiologique
Une ANOVA de type 3 a montré un effet principal significatif de la source de fixation de l'objectif sur le rapport VRC LF/HF après contrôle de l'engagement dispositionnel, $F(2, 88) = 2,80$, $p = 0,065$, $\eta p2 = 0,06$. Des régressions linéaires par paires post-hoc ont révélé que l'objectif autodéfini entraînait un rapport LF/HF plus faible que l'absence d'objectif ($t = -2,31$, $p = 0,070$). Cependant, aucune différence n'a été observée entre les objectifs autodéfinis et les objectifs assignés ($t = -0,55$, $p = 0,584$) ou entre les objectifs assignés et l'absence d'objectif ($t = -1,79$, $p = 0,155$).

### 4.2.2 Engagement émotionnel (état)
#### 4.2.2.1 Auto-évaluation
Une ANOVA de type 3 a montré un effet principal significatif de la source de l'objectif sur l'activation autorapportée après contrôle de l'engagement dispositionnel, $F(2, 91) = 26,14$, $p < 0,001$, $\eta p2 = 0,36$. Des régressions linéaires par paires post-hoc ont révélé que l'activation autorapportée était plus élevée lorsque les objectifs étaient autoassignés comparé à la situation sans objectif ($t = 6,40$, $p < 0,001$). De même, l'activation autorapportée était plus élevée lorsque l'objectif était assigné par rapport à l'absence d'objectif ($t = 5,72$, $p < 0,001$). Cependant, aucune différence n'a été observée entre les objectifs autodéfinis et assignés ($t = 0,80$, $p = 0,425$).

Une ANOVA de type 3 a montré un effet principal significatif de la source de fixation de l'objectif sur la valence autorapportée après contrôle de l'engagement dispositionnel, $F(2, 91) = 6,70$, $p = 0,002$, $\eta p2 = 0,13$. Des régressions linéaires par paires post-hoc ont révélé que la valence autorapportée était plus élevée lorsque les objectifs étaient autodéfinis comparé à aucun objectif ($t = 3,34$, $p = 0,004$). De même, la valence autorapportée était plus élevée lorsque l'objectif était assigné par rapport à l'absence d'objectif ($t = 2,93$, $p = 0,008$). Cependant, aucune différence n'a été observée entre les objectifs autodéfinis et assignés ($t = 0,47$, $p = 0,639$).

#### 4.2.2.2 Physiologique
Une ANOVA de type 3 n'a montré aucun effet principal significatif de la source de fixation de l'objectif sur la fréquence respiratoire après avoir contrôlé l'engagement dispositionnel, $F(2, 84) = 1,50$, $p = 0,230$.

### 4.2.3 Engagement comportemental
#### 4.2.3.1 Auto-évaluation
Une ANOVA de type 3 a montré un effet principal significatif de la source des objectifs sur la vigueur autorapportée après contrôle de l'engagement dispositionnel, $F(2, 91) = 5,64$, $p = 0,005$, $\eta p2 = 0,11$. Des régressions linéaires par paire post-hoc ont révélé que la vigueur autorapportée était plus élevée lorsque les objectifs étaient autodéfinis que lorsqu'il n'y avait pas d'objectif ($t = 2,46$, $p = 0,032$). De même, la vigueur autorapportée était plus élevée lorsque l'objectif était assigné par rapport à l'absence d'objectif ($t = 3,30$, $p = 0,004$). Cependant, aucune différence n'a été observée entre les objectifs autodéfinis et les objectifs assignés ($t = -0,81$, $p = 0,418$).

#### 4.2.3.2 Physiologique
Une ANOVA de type 3 a montré un effet principal significatif de la source de fixation de l'objectif sur l'intensité de l'effort physique après contrôle de l'engagement dispositionnel, $F(2, 88) = 9,47$, $p < 0,001$, $\eta p2 = 0,18$. Des régressions linéaires par paires post-hoc ont révélé que l'intensité de l'effort physique était significativement plus élevée lorsque l'objectif était autodéfini par rapport à l'absence d'objectif ($t = 4,08$, $p < 0,001$). De même, l'intensité de l'effort physique était plus élevée lorsque le but était assigné par rapport à l'absence de but ($t = 3,16$, $p = 0,004$). Cependant, nous n'avons pas observé de différence lorsque l'objectif était autodéfini par rapport à un objectif assigné ($t = 0,95$, $p = 0,345$).

*4.3 H3 : Les objectifs choisis par soi-même amèneront à une meilleure performance comparativement aux objectifs assignés ou l'absence d'objectifs.*

Une ANOVA de type 3 a montré un effet principal significatif de la source de fixation de l'objectif sur le temps de performance après contrôle de l'engagement dispositionnel, $F(2, 90) = 34.83$, $p < .001$, $\eta p2 = .44$. Les régressions linéaires par paires post-hoc ont révélé que la performance était meilleure (temps inférieur) lorsque les objectifs étaient autodéfinis, comparé aux objectifs assignés ($t = -2,28$, $p = 0,025$) et à l'absence d'objectif ($t = 7,900$, $p < 0,001$). De même, la performance était meilleure lorsque le but était assigné par rapport à l'absence de but ($t = 5,80$, $p < 0,001$). La Figure 3 illustre ces résultats.

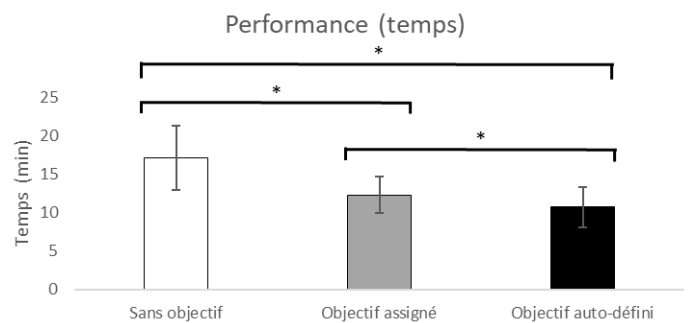

**Figure 3. Performance (temps) par condition**

## 5 DISCUSSION

L'étude actuelle a examiné l'effet de la source de fixation d'objectifs (autodéfinis, assignés, sans objectif) sur l'autonomie, l'engagement et la performance des participants. Les résultats indiquent que la perception de l'autonomie des participants était significativement plus élevée lorsque les objectifs étaient fixés par eux-mêmes que lorsqu'ils étaient assignés. Ceci soutient l'hypothèse H1. Le fait de choisir leurs propres objectifs a permis aux participants d'avoir une certaine latitude décisionnelle et un certain contrôle dans la tâche. En revanche, le fait d'avoir un objectif imposé par une source externe a conduit les participants à se sentir contrôlés. Une plus grande autonomie sur le lieu de travail est synonyme d'une plus grande motivation autodéterminée. Dans l'optique de la TAD, ce résultat implique que **les objectifs autodéfinis conduiront à des résultats plus positifs à long terme en termes de rétention des employés, d'acceptation de la technologie, de bien-être, de comportement de citoyenneté organisationnelle et de performance** (Deci et al., 2017 ; Meyer & Gagne, 2008 ; Venkatesh et al., 2002). Ces résultats contribuent également à la théorie de la fixation des objectifs. Bien que cette théorie stipule que les objectifs autodéfinis semblent être plus efficaces pour la poursuite à long terme des objectifs, les mécanismes qui sous-tendent ce résultat sont peu étudiés dans un contexte de tâches répétitives. Les résultats actuels suggèrent qu'une perception accrue de l'autonomie peut être l'un des mécanismes par lesquels les objectifs fixés de type autodéfinis conduisent à une meilleure poursuite sur le long terme par rapport aux cas où les objectifs sont assignés. Les recherches futures devraient tester spécifiquement cette hypothèse. Aussi, il faudrait tester la relation entre l'autonomie perçue et l'internalisation des buts (cf 2.2). Selon la théorie de la fixation des objectifs, plus d'autonomie pourrait amener à une meilleure internalisation des buts et donc une meilleure poursuite d'objectif à long terme.

En termes d'engagement, **nos résultats montrent que les participants étaient plus engagés lorsqu'un objectif était présent (qu'il soit autodéfini ou assigné) que lorsqu'aucun objectif n'était présent.** Cependant, l'engagement n'était pas différent entre les conditions d'objectifs autodéfinis et assignés. L'hypothèse H2 n'a donc pas été confirmée. En termes de temps de performance, **les objectifs autodéfinis ont conduit à une meilleure performance (temps inférieur) par rapport à l'objectif assigné et à l'absence d'objectif**. Ceci soutient l'hypothèse H3. La constance de l'engagement, quelle que soit la condition de l'objectif fixé par l'individu par rapport au cas où l'objectif est assigné, est en contradiction avec le modèle théorique de la TAD. En effet, cette dernière stipule qu'une satisfaction accrue de l'autonomie entraîne un engagement accru et donc une meilleure performance. Nos résultats suggèrent donc que l'engagement n'explique pas la relation entre une autonomie accrue et une performance accrue. D'autres facteurs, tels que la volonté de fournir des efforts, l'internalisation de l'objectif ou l'engagement envers l'objectif (goal commitment), peuvent plutôt expliquer cette relation (Locke & Latham, 2002). Les recherches futures devraient tester ces relations.

Néanmoins, **nos résultats,** qui sont en ligne avec un étude précédente (Passalacqua, Léger, et al., 2020), **impliquent que les objectifs fixés par les employés eux-mêmes mènent à de meilleurs résultats que lorsque les objectifs sont assignés ou en absence d'objectifs**. Lors de la gamification d'un système avec fixation d'objectifs, les concepteurs et les gestionnaires devraient laisser les employés choisir leur propre objectif plutôt que d'en assigner un. **La gamification dans laquelle l'employé participe activement à la définition de ses propres objectifs a les meilleures chances de produire une motivation plus autodéterminée, ce qui a un impact à long terme sur de multiples facettes de la performance et du bien-être des employés et de l'entreprise**.

Cette étude présente deux limites. Notre étude a mesuré toutes les variables dépendantes à court terme. Par conséquent, nous avons utilisé les cadres théoriques de la TAD et de la théorie de la fixation des objectifs pour extrapoler sur le long terme. De plus, notre étude n'a pas mesuré certaines variables qui auraient pu expliquer la relation entre l'autonomie et la performance, c'est-à-dire l'engagement envers les objectifs (goal commitment), la volonté de fournir des efforts et l'internalisation des objectifs.

# 6 Références